\documentclass[12pt]{article}
\begin{document}
\title{Chirality and Cosmic Origins of Life}
\author{B.G. Sidharth\\
International Institute for Applicable Mathematics \& Information Sciences\\
Hyderabad (India) \& Udine (Italy)\\
B.M. Birla Science Centre, Adarsh Nagar, Hyderabad - 500 063
(India)}
\date{}
\maketitle
\begin{abstract}
Chirality or the property that distinguishes lefthandedness from
righthandedness is an important aspect of the universe, starting
from neutrinos, which are lefthanded. Fifteen years ago the author
had proposed that life on the earth was formed through a dual
process -- several key ingredients being transported from outer
space to the earth by comets or meteorites and these in turn
interacting with compounds already cooked up in the earth's seas.
Several recent observations point to the fact that the amino acids
brought out to the earth by meteorites are lefthanded, as in
terrestrial life forms. Experiments in the laboratory however throw
up equal numbers of lefthanded and righthanded amino acids, what are
called racemic mixtures. Not only would the latest observations
endorse this dual mode theory of the origin of life, but on the
other hand it would point to the key trigger for life processes
itself. This is because racemic mixtures are not optically active,
unlike the handed or chiral constituents. It is this activity of
harnessing solar radiation and converting it into chemical energy,
and storing this in the sugar bonds, as in photosynthesis, that
characterizes life.
\end{abstract}
\section{Introduction}
Recently there have been some renewed claims that ready made life
reached the earth from outer space \cite{bgs}. For example, R.Joseph
and others try to make an elaborate  case for life having been
transported to the earth from outer space.These conclusions are at
best very speculative and very debatable, if not dubious.\\
The  contention is that life was brought to earth via comets or
asteroids, a view propounded by, amongst others Chandra
Wickramasinghe and Sir Fred Hoyle. The original idea itself is more
than a century old, having been put forward by Arrhenius and others.
These ideas went by the name Lithospanspermia. The argument is that
microbes could be cold stored as spores and could be brought back to
life millions of years hence. All this is within the realm of
possibility, but in the absence of unambiguous proof it cannot be
treated as being plausible. Undoubtedly, debris produced by
supernovae could have bombarded the earth. Footprint on the earth
could be the iron and Carbon 12 in ancient rocks that are over four
billion years old. But from here to conclude that what was brought
down to earth was readymade life itself is a very far cry.\\
For example let us take the case of the ALH 84001 meteorite which
was supposedly ejected from Mars a few billion years ago and fell in
the Antarctic region a few thousand years ago. In the mid nineties
this meteorite shot into prominence because of claims that it had
distinct traces of fossilized micro life, as claimed by Dr. David
Mckay and a few others. This claim has been contested because
similar shapes have been found on terrestrial samples as well. In
fact it has been pointed out that these could well be mineral grain
deposits. Similar arguments can be put forward in the case of the
other carbonaceous chondrites.\\
On the other hand, the prevailing view that the first organisms were
entirely cooked up in the earth's soup may also be far fetched.
Francis Crick once noted how improbable this would be. The author
suggested in the early nineties that a dual mode origin of life was
more likely. That is key ingredients like amino acids, but not yet
fully formed life had reached the earth from outer space and
chemically interacted with other ingredients present on the earth,
to form life \cite{bgs,bgs2,bgs3}.
\section{A dual mode origin}
The basis for these arguments was the observation that several
complex molecules have been discovered in interstellar space, for
example, in the cool dust clouds of the Orion Nebula and in the
constellation of Sagittarius. Observations with telescopes,
spectroscopes, radio telescopes and even orbiting observatories have
confirmed the presence of molecules like methyl cyanide, water
vapour, formaldehyde, methyl alcohol and even the potable ethyl
alcohol. Clearly there are certain organic molecules in the cool
dust clouds spread across outer space.\\
Over one hundred and twenty molecules including those which chemists
designate as Methanol and Ethanol, from diatomic molecules through
those containing thirteen atoms have been detected in the dark
interstellar clouds which are opaque to light. These appear to be
nothing less than factories for building complex molecules. The
mechanism could be that some of the simpler molecules are frozen in
ice droplets which are bombarded by ultra violet radiation from very
near by young stars, cosmic rays and ions, all these inducing the
formation of very complex molecules by the process of breaking up of
molecules and recombination. Interestingly this has been confirmed
by simulations on the earth. The polycyclic aromatic hydrocarbons
(of aerosol spray and global warming notoriety) under interstellar
conditions convert to complex molecules, like alcohols, ethers and
quinons. These are ubiquitous in living organisms today, helping in
various energy transfer processes like photosynthesis and the
ability to absorb ultra violet radiation which is harmful to, for
example amino acids. The universe exhibit traces of polycyclic
aromatic hydrocarbons. These are the most abundant class of carbons
in the universe and may be containing about twenty percent of all
the carbon.\\
This apart the space crafts Giotto and Vega which flew by Comet
Halley glimpsed carbon rich molecules while space based observations
revealed the presence of Ethane and Methane in Comets Hyakutake and
Hale Bopp. Space dust reveals organic carbon. Interestingly some
thirty tons of such carbon is brought down to the earth each day by
the interstellar dust. Meteorites have shown nucleo basis, ketones,
quinines, carboxylic acids, amines and amides. In fact as many as
eight of the twenty amino acids involved in life processes have been
identified besides some sixty others. This August, NASA announced
that an analysis of data from its Stardust mission revealed, for the
first time the presence of the amino acid glycine in an icy comet.\\
Comets which inhabit the cool and dark regions of the solar system
definitely contain the building blocks of life. But again, it would
be a giant leap of faith, if we say that they contain living
organisms. There is increasing evidence through spectroscopic, space
craft and even laboratory examinations of debris fallen on the
earth, to show that the frozen dirty ice balls, as comets have been
characterized, contain not just molecular compounds of carbon,
hydrogen, nitrogen and oxygen, but also even more complicated sugar
related substances. Studies at NASA's Ames Research Centre indicate
the presence of polyhydroxilated compounds as well. Studies at the
Russian Academy of Sciences have confirmed the possibility of
abiogenous synthesis of complex organic compounds (monomeric units
of nucleic acids) on the surface of comets, asteroids, meteorites
and space dust particles in outer space.\\
The presence of sugar related compounds and other complex molecules
in meteoritic and cometary objects will support the author's theory
of dual origin. It is important to keep the background in mind. Some
of these molecules, possibly proteins or amino acids-but not yet
living organisms-are very likely to have been transported to earth,
and further biochemical reactions would have taken place on the
earth itself, with for example fats. Such a hybrid view for the
origin of life is consistent with observations and experiments.\\
There are intriguing footprints of outer space on earth based living
organisms. The all important amino acids in nature come as left
handed molecules and also right handed molecules reminiscent of a
right handed spiral conch shell and a left handed spiral conch. The
amino acids produced in the laboratory like the Urey-Miller type
experiments show equal quantities of the left handed and right
handed varieties, which is reasonable. However in life processes,
the left handed molecules predominate over the right handed
molecules. Interestingly in the amino acids found on meteorites, we
have exactly this preponderance of  left handed amino acid
molecules!
\section{The Key Trigger}
How can extra terrestrial molecules trigger off chemical reactions
leading to the origin of life? Firstly complex molecules containing
six or more carbon atoms are known to produce amino acids in warm
acidic water, for example on the earth. This apart molecules with as
many as fifteen carbon bonds were also created. These molecules form
droplet-like capsules. Indeed such capsules, the precursors of cell
walls were also exhibited by extracts from meteorites-when organic
compounds from these meteorites were mixed with water, they
assembled into cell membrane like structures with complex organic
compounds. Actually what happens is that these molecules are
amphidilic: One of their ends has a preference for water, while the
other behaves in an exactly opposite manner. The ends which prefer
water form the outer circles of the cell like structures while at
the same time hiding within are the inner ends. Such structures
could also house other interstellar components, for example quinons
which could harness light and other forms of energy required for
life processes.\\
It appears that amino acids, quinons, amphibilic molecules and the
like were transported to the earth by meteoritic dust or cometary
fragments. These could well have kick started the first life
processes on the earth. But there could be a key trigger as we will
see now.\\
On the face of it, the universe appears to be symmetric, including
the left-right symmetry. However a breaking of a symmetry is
crucial, as is well known. Thus the neutrinos are lefthanded, that
is display chirality or handedness. This refers to the crucial
difference between the lefthanded and the righthanded -- though
mirror images, they cannot be exactly superimposed on each other.\\
This chirality carries over into the realm of molecules. Such a
molecule displaying handedness like the neutrino is called
enantiomer. A mixture of equal amounts of two enantiomers, that is
lefthanded and righthanded molecules is said to be racemic. The
crucial fact is that racemic, that is equally handed mixtures are
not optically active. On the other hand separately the enantiomer
constituents are optically active.\\
Optical activity itself arises from the interaction of chiral
molecules with polarized light -- they can rotate the plane of
polarization. Indeed such optical activity carries over to other
regions of the spectrum, for example in the microwave region.\\
It is worth noting that the differently handed constituents of
chiral compounds have different properties, an interesting example
being that of righthanded sugar (dextrose) and its lefthanded
counterpart. Ultimately the differences are due to the chirality
inherent in biological system.\\
As we saw above, the amino acids are chiral and so too are the
sugars. It must also be noted that the chirality of life on earth
could well be an accident -- life on other planets could well have
the opposite chirality. What is important is that the mixture of
amino acids, which are the building blocks of proteins should not be
racemic. This would ensure optical activity in a self-organization context.\\
Let us consider the above points in a little greater detail. As far
as optical activity is concerned, we could think  of a racemic
mixture as a combination of two polarizers, turned perpendicular to
one another, light passing through one is polarized in the $x$
direction let us say. When this passes through the second polarizer
this rotated ninety degrees with respect to the first one, then it
is completely absorbed.\\
With regard to self organization, we note that there is a free
interplay of solar radiation with the original organic molecules, so
that we have to consider open systems and far from equilibrium
situations. Let us first start with crystals which to a certain
extent mimick life processes. As is well known crystals are ordered
arrangements of molecules or atoms which nucleate at certain sites.
They can spontaneously grow, and this is a process where the
molecules are interacting with one another through forces like Van
der Waal interaction. They form bonds, lowering the potential energy
in the process. Though ordered, crystals are an example of an
equilibrium thermodynamic process. Let us now consider a process of
ordering with activity unlike inert crystal. A good example of this
is that of the Benard convection cells \cite{prig1}. As is well
known, to realize these cells, let us consider water that is
stationary between two plates. When the difference of the
temperature between the plates crosses a critical value, highly
organized Benard cells are seen to form. However it must be observed
that in this case the system is open, in that it receives heat from
outside, facilitating the formation of the so called dissipative
structures, unlike the equilibrium structures such as crystals,
while crystals were a phenomena pertaining to the intermolecular
interaction. In this latter case external conditions drive the
system away from equilibrium, more quantitatively, the affinity
parameter $A$ vanishes for the equilibrium situation, whereas an
increase in its magnitude reflects a departure from the equilibrium
situation to the non equilibrium regime.\\
Here in are the seeds of self organization. A key role in these
processes is played by auto catalysis leading to a multiplication of
certain molecules, as is well known.\\
There is a further novelty in these far from equilibrium structures,
brought out most simply by the well known chemical clock resulting
from the $B-Z$ reaction. Here we see that the colours of certain
regions change at very regular time intervals. The point here is
that the changing of the colours is a bulk phenomenon affecting
millions of molecules, which seem to act in unison. In other words
there is a new feature, viz., a communication between the
molecules.\\
Let us consider a simple mathematical model in this regard -- a
random and indefinitely long sequence of a finite number of symbols
like $0,1$ or $A,B,C$, without any constraint on the number of times
a symbol can appear in this indefinite sequence. It is always
possible to abstract a sub sequence $S$ which repeats itself
elsewhere along the line at least once. This would then happen not
just twice but several times. Clearly there would be other sub
sequences like $S$ which would also exhibit this behaviour, though
one of them may have a greater frequency than the others. Such
patterns are always possible in an open system, for example given
enough time, energy and material. This recurrent pattern resembles
replication. To idealize the situation we could say that if there
were $N$ such sequences of a type, then in an average time $\tau$
the increase of this number $\Delta N$ can be modeled by
\begin{equation}
\frac{dN}{dt} = \frac{\Delta N}{\tau} = \beta N^m\label{A}
\end{equation}
where, if $\tau$ is sufficiently small compared to the time scales,
we could consider the time derivative of $N$ as shown in (\ref{A}),
which leads to
\begin{equation}
N^{(1-m)} = (1-m) \beta T\label{2}
\end{equation}
where $T$ is the total time that is under consideration. For
instance in a typical Random Walk type of a scenario $m$ would be
half. In any case we can see here the replication and the beginning
of communication in the sense of the chemical clock above,
remembering that poly nucleotide sequences stored information, given
the right conditions. In our mathematical example, the specificity
of each of the $N$ sub sequences is itself information.The rest is
junk (DNA). Another point to bear in mind is that the conservation
of information in polymer sequences distinguishes biological self
organization from
the chemical counterpart \cite{schuster}.\\
Ultimately it is the energy of solar radiation on the earth which is
converted into chemical energy, and is stored in the bonds of sugar.
This is done in algae and plants through the process of
photosynthesis.\\
In recent years, dramatic evidence for this dual mode origin of life
has been coming in from different studies. The recent work of
Pizzarello in Arizona State University confirms the above scenario.
The work of Prof. Ronald Breslow of Columbia University, New York
and his coworker M. Levine, for another example. They have found
that the predominant left handed amino acids present in the life
forms could indeed have been delivered by meteorites from deep outer
space, thus seeding life on the earth. This new evidence was found
on the surfaces of meteorites which had crashed into the earth in
relatively recent times. Dr. Breslow exposed amino acid chemical
precursors to the amino acids found on the meteorites and found that
these cosmic molecules could transfer their left handedness to the
simple amino acids found in life forms. According to Breslow this is
the mechanism by which there is an excess of left handed amino acids
in life-even a small excess to begin with could grow in numbers.\\
Another dramatic confirmation has come in. An important component of
early genetic material found in meteorite fragments is
extraterrestrial in origin, say scientists from Europe and the USA.
They say their research in Earth and Planetary Science Letters
provides evidence that life's raw materials came from sources beyond
the Earth.\\
The Materials they have found include the molecules uracil and
xanthine, which are precursors to the molecules that make up DNA and
RNA, and are known as nucleobases. The team discovered the molecules
in rock fragments of the Murchison meteorite, which crashed in
Australia in 1969.\\
They tested the meteorite material to determine whether the
molecules came from it were a result of contamination when the
meteorite landed on Earth. The analysis showed nucleobases contain a
heavy form of carbon which could only have been formed in
space-those formed on Earth consist of a lighter variety of
carbon.\\
Lead author Dr. Zita Martins, of the Department of Earth Science and
Engineering at London \cite{martin} says:\\
"We believe early life may have adopted nucleobases from meteoritic
fragments for …. which enabled them to pass on their successful
features to subsequent generations.\\
Between 3.8 to 4.5 billion years ago large numbers of rocks similar
to the Murchison down on Earth at the time when primitive life was
forming. The heavy bombardment large amounts of meteorite material
to the surface on planets like Earth and Mars.\\
Co-author Professor Mark Sephton, also of Imperial's Department of
Earth Science: believes this research is an important step in
understanding how early life might have…\\
"Because meteorites represent left over materials from the formation
of the solar system components for life-including nucleobases-could
be widespread in the cosmos. … life's raw materials are discovered
in objects from space, the possibility of life spring… right
chemistry is present becomes more likely."\\
Even more evidence was unveiled early this year \cite{barnun} by
Professor Akiva Bar-Nun of the Department of Geophysics and
Planetary Sciences at Tel Aviv University. He and coworkers found
independently that key missing ingredients for life to form in the
earth's ancient environ, were available in comets. . "When comets
slammed into the Earth through the atmosphere about four billion
years ago, they delivered a payload of organic materials to the
young Earth, adding materials that combined with Earth's own large
reservoir of organics and led to the emergence of life," says Prof.
Bar-Nun. Specifically Prof. Bar-Nun's work dealt with some of the
noble gases in the icy comets.\\
Incidentally, all this means that life would be well spread out in
the universe and is not unique to the earth-it could well have
formed a few billion years ago on Mars, for instance.

\end{document}